\documentclass[intlimits,twoside,a4paper]{article}

\usepackage[cp1251]{inputenc}
\usepackage{amssymb}
\usepackage{amsmath}
\usepackage[usenames]{color}

\usepackage{epsfig}

\usepackage[eqsecnum]{cmpj3}

\issue{2019}{22}{3}{33501}
\doinumber{10.5488/CMP.22.33501}

\title%
{Short- and long-range contributions to equilibrium and transport properties of solid electrolytes}
\author[G.~Bokun \textsl{et al.}]{G.~Bokun\refaddr{label1}, I.~Kravtsiv\refaddr{label2},  M.~Holovko\refaddr{label2}, V.~Vikhrenko\refaddr{label1}, D.~di~Caprio\refaddr{label3}}
\addresses{
	\addr{label1}Belarusian  State Technological University, 13a Sverdlov St., 220006 Minsk, Belarus
	\addr{label2} Institute for Condensed Matter Physics of the National Academy of Sciences of Ukraine, \\
	1 Svientsitskii St., 79011 Lviv, Ukraine
	\addr{label3} Institute of Research of Chimie Paris, CNRS-Chimie ParisTech, 11 rue Piere et Marie Curie, \\ 75005 Paris, France
}

\date{Received	May 30, 2019, in final form June 26, 2019}

\begin{document}

\maketitle

\begin{abstract}
	
Condensed ionic systems are described in the framework of a combined approach that takes into account both long-range and short-range interactions. Short-range interaction is expressed in terms of mean potentials and long-range interaction is considered in terms of screening potentials. A system of integral equations for these potentials is constructed based on the condition of the best agreement of the system of study with the reference system. In contrast to the description of media with short-range interactions, in this text the reference distribution includes not only the field of mean potentials but also Coulomb interaction between particles. A one-component system made of ions in a neutralizing background of fixed counterions is considered. The model can be used to describe solid ionic conductors. In order to study the movement of cations on the sites of their sub-lattice, the lattice approximation of the theory is employed based on the calculation of the pair distribution function. Using the collective variables approach, a technique improving the starting expression for this function is proposed. Notably, the neglected terms in the expansion of this function are approximated by single-particle terms ensuring that the normalization condition is satisfied. As a result, the applicability of the theory is extended to a wide region of thermodynamic parameters. The chemical potential and the diffusion coefficient are calculated showing the possibility of phase transitions characteristic of the system.                  	

\pacs 5.20.-y, 61.72jd, 64.30.-t, 65.40.-b, 66.10-x
\end{abstract}

\section{Introduction}

The problem of combined action of long-range electrostatic and short-range van der Waals interactions has a long history. Its importance was demonstrated in the study of stability of colloids and disperse systems \cite{Derjaguin39,Derjaguin39a,Derjaguin41,Derjaguin41a,Verwey48}. In the last decades the interest for these interactions was renewed due to investigations of complicated behavior of nanoscale and biological objects that can be described by the competing interactions \cite{Sear99,Stradner04,Rozovsky05,Campbell05,Imperio06,Ciach10,Almarza16,Rauha17}.

For simpler ionic systems, no more than short-range steric interactions were initially taken into account in the theory of electric double layers \cite{Stern24,Henderson83,Borukhov97,Caprio04}. Later the short-range hard-core repulsion was included based on statistical-mechanical integral equation theories while electrostatic interactions were accounted for in the framework of the mean field approximation \cite{Henderson83,Blum75,Blum77,Yukhnovskii80,Golovko85,Hummer93,Bhuiyan09}.

Recently, the interest for competition of short- and long-range interactions in ionic systems was renewed due to an important role played by ionic liquids in industrial applications. Extremely high ion concentrations in ionic liquids strongly increase the importance of short-range interactions and correlations. It was shown that with accounting of the short-range interactions on the mean field level in addition to the Coulomb ones the voltage dependence of the differential capacitance becomes closer to experimental findings \cite{Goodwin17,Yin18}. An important influence of a competition of short-range dispersion (or van der Waals) and long-range Coulomb interactions on phase transitions in a lattice model of ionic systems was demonstrated in \cite{Ciach01}. The short-range dispersion interactions alongside with ionophobicity or ionophilicity of electrodes can lead to a variety of voltage dependences of electrical double layer capacitance in ionic liquid-solvent mixtures \cite{Cruz19}. Short-range interactions and correlations up to the third neighbors were taken into account in \cite{Bokun18,Patsahan19} based on the method of collective variables \cite{Yukhnovskii80,Golovko85}. Significant influence of short-range interactions and interparticle correlations on the charge distribution in the near electrode region and in the bulk was observed \cite{Patsahan19}. The variation of the host crystal field in the vicinity of the electrode that is analogous to its ionophobicity or ionophilicity can lead to an anomalous behavior of the electric double layer capacitance \cite{Bokun18}. In fact, attractive short-range van der Waals interactions being opposite to electrostatic repulsive ones make ionic systems similar to systems with competing interactions \cite{Sear99,Stradner04,Rozovsky05,Campbell05,Imperio06,Ciach10,Almarza16} that have been widely studied in recent years. 

Solid electrolytes represent a particular type of ionic systems that are widely used in electric current sources, energy storage systems, etc. \cite{Arunkumar12,Ulihin13,Melchionna14,Cao14}. In many cases, they can be considered as one-component mobile charges moving on the potential relief of the host system, which play the role of neutralizing background as well \cite{Kroger64,Knauth02,Kharton04}.   

In this article we present a statistical mechanical approach that allows us to account for the combined action of short- and long-range interactions in the system. The free energy functional of a condensed system \cite{Bokun18a} is generalized to include the presence of long-range interactions. Notably, the approach proposed in \cite{Rott79, Rott75} and generalized to non-homogeneous systems in \cite{Narkevich78, Narkevich82}  is improved by considering the reference distribution that consists of not only single-particle cell potentials but also of the long-range part of inter-molecular interactions. The latter can be calculated in the framework of the collective variables method \cite{Yukhnovskii80,Golovko85}. As a result, a joint approach combining the method of mean potentials and the collective variables formalism is developed. This approach makes it possible to take into account the contributions from short-range as well as from long-range interactions, both of which are important in the study of condensed states of matter. Among other things, this allows one to describe the emergence of competing interactions, which have been widely employed in the study of inhomogeneous systems \cite{Sear99,Stradner04,Rozovsky05,Campbell05,Imperio06,Ciach10,Almarza16,Ciach01}.

\section{Distribution functions of the reference system}

Our research employs lattice description, whereby the volume $V$ of the system, occupied by $N$ particles, is divided into $M$ cells $\omega_i$ ($i=1\ldots M$). In general, the cells are not identical in size and each cell can be either vacant or occupied by one particle. We introduce $\alpha_i\equiv i_0=0 \text { or } i_1=1$ as the occupation number of cell $i$. Accordingly, we can write
\begin{align}
\sum_{i=1}^{M}\alpha_i=N,\qquad \sum_{i=1}^{M}\omega_i=V,\qquad c=\frac{N}{M}\,, \qquad \omega=\frac{V}{M}\,, 
\end{align}     
\noindent where $c$ is the average concentration of particles, $\omega$ is the average volume of a cell. The distribution of concentrations is, therefore, determined by the quantities $c_i$. It is convenient to view vacant cells as ``virtual'' particles neither interacting with each other nor with the real particles. The position of each particle is, therefore, given by a vector $q_{\alpha_i}$, where $q_{1_i}$ denotes the coordinate of a real particle and $q_{0_i}$ denotes that of a virtual particle. A particular configuration of a system is characterized by a set of parameters $\alpha_i$ and the respective coordinates $q_{\alpha_i}$. The energy of the system corresponding to a particular configuration can therefore be written as follows: 
\begin{align}
\label{ham}
H_M=\frac{1}{2}\sum_{i=1}^{M}\sum_{j=i}^{M}\left[\Phi(q_{\alpha_i}q_{\alpha_j})+V(q_{\alpha_i}q_{\alpha_j})\right],
\end{align}   
where $\Phi(q_{\alpha_i}q_{\alpha_j})$ is the short-range pair potential, $V(q_{\alpha_i}q_{\alpha_j})$ is the long-range part of interaction, and
\begin{align}
\Phi(q_{0_i}q_{\alpha_j})=0\,,\qquad V(q_{0_i}q_{\alpha_j})=0.
\end{align}

Thus, the configurational integral of the system is as follows:
\begin{align}
\label{conf}
Q_N &= \frac{1}{\omega^{M-N}}\sum_{\alpha_1=0}^{1}
\sum_{\alpha_2=0}^{1}\ldots \sum_{\alpha_i=0}^{1}\ldots \sum_{\alpha_M=0}^{1}\int_{\omega_1} \rd q_{\alpha_1}
\int_{\omega_2} \rd q_{\alpha_2}\ldots \int_{\omega_i} \rd q_{\alpha_i}\ldots \nonumber \\
&\ldots \int_{\omega_M}\rd q_{\alpha_M}\exp\left[-\beta H_M\left(\left\{q_{\alpha_i},\alpha_i\right\}\right)\right].
\end{align}

In what follows we consider the case where the volumes of all the cells are equal ($\omega_i=\omega$).

In order to calculate the configurational integral, we expand expression (\ref{conf}) to the states of the reference system, which, similar to \cite{Bokun18a}, will include the long-range interaction energy in addition to the mean potentials.

The Hamiltonian of the reference system is
\begin{align}
\label{H0}
H_0=\frac{1}{2}\sum_{1}^{M}\sum_{j(i)}^{M}V(q_{\alpha_i},q_{\alpha_j})+\sum_{i=1}^{M}U(q_{\alpha_i})\,,
\end{align}  
where $U(q_{\alpha_i})$ is the energy of self-consistent field, which is expressed in terms of the mean potentials as follows:
\begin{align}
\label{uu}
U(q_{\alpha_i})=\sum_{j(i)}^{M}\phi_j(q_{\alpha_i})\,,
\end{align}
where $\phi_j(q_{\alpha_i})$ is the mean potential acting from a particle in cell $j$ on a particle located at $q_{\alpha_i}$. Due to the definition of the quantity $\phi_j(q_{\alpha_i})$, we can write the singlet distribution function of the reference system as follows:
\begin{align}
\label{f1}
F_1^0(q_{\alpha_i})=c_{\alpha_i}\frac{\exp\Big[-\beta\sum\limits_{j(i)}^{M}\phi_j(q_{\alpha_i})\Big]}{Q_{\alpha_i}}\,,
\end{align} 
where 
\begin{align}
\label{qq}
Q_{\alpha_i}=\int_{\omega_i}\exp\Big[-\beta\sum_{j(i)}^{M}\phi_j(q_{\alpha_i})\Big] \rd q_{\alpha_i}\,.
\end{align}

In expressions (\ref{f1})--(\ref{qq}) we account for the fact that the singlet distribution function should be normalized to give the concentration of particles of the ``sort'' $\alpha_i$ such that the quantities $c_{1_i}=c_i$, $c_{0_i}=1-c_i$ denote the concentrations of particles and vacancies in cell $i$.  

Similar to \cite{Bokun18a}, when we expand (\ref{conf}) to the states of the reference system, we take into account only the first two terms, for which it is sufficient to know the singlet function (\ref{f1}) and the pair distribution function. According to \cite{Yukhnovskii80,Golovko85}, the latter can be presented in the form 
\begin{align}
\label{f22}
F_2^0(q_{\alpha_i},q_{\alpha_j})=F_1^0(q_{\alpha_i})F_1^0(q_{\alpha_j})
\exp\left[-\beta\sigma(q_{\alpha_i},q_{\alpha_j})\right]\left(1+ \ldots\right),
\end{align} 
where $\sigma(q_{\alpha_i},q_{\alpha_j})$ is the screening potential. 

The pair distribution function should satisfy the normalization condition
\begin{align}
\label{f10}
F_1^0(q_{\alpha_i})=\sum_{\alpha_j=0}^{1}\int_{\omega_j}F_2^0(q_{\alpha_i},q_{\alpha_j}) \rd q_{\alpha_j}.
\end{align} 

Taking into account only the terms (\ref{f22}) and presenting the contribution from the rest of the terms in a factorized form, we can write the pair distribution function of the reference system herein as follows:
\begin{align}
\label{f23}
F_2^0(q_{\alpha_i},q_{\alpha_j})=F_1^0(q_{\alpha_i})F_1^0(q_{\alpha_j})S_j^{-1}(q_{\alpha_i})S_i^{-1}(q_{\alpha_j})
\exp\left[-\beta\sigma(q_{\alpha_i},q_{\alpha_j})\right].
\end{align} 

The additional factors $S_j^{-1}(q_{\alpha_i})$ are introduced to satisfy the normalization condition. Replacing expressions (\ref{f1}), (\ref{qq}), (\ref{f23}) into equation (\ref{f10}) and due to equation (\ref{uu}), we obtain 
\begin{align}
\label{s}
S_j(q_{\alpha_i})=\sum_{\alpha_j=0}^{1}\frac{c_{\alpha_j}}{Q_{\alpha_j}}\int_{\omega_j}
\frac{1}{S_i(q_{\alpha_j})}\exp\left[-\beta\sigma(q_{\alpha_i},q_{\alpha_j})+U(q_{\alpha_j})\right]
\rd q_{\alpha_i}\,.
\end{align}

The factors defined by expression (\ref{s}) ensure the correct normalization and, hence, the correct asymptotic behavior of the function (\ref{f10}). Relationships (\ref{f23}), (\ref{s}) should be closed by a system of equations that define mean potentials.

\section{The single-particle cell potentials}

The expression for the configurational integral does not contain mean potentials. By adding and subtracting these potentials to and from the Hamiltonian of the reference system, we can present the N-particle Gibbs function as the product of the Gibbs function of the reference system and the factor that can be calculated using the perturbation theory. As a result, after the expansion on cumulants we obtain the sum every term of which depends on the potentials $\phi_j(q_{\alpha_i})$ while the total sum remains invariant with respect to the choice of these potentials. This fact allows us to develop a scheme of self-consistent calculation of potentials \cite{Bokun18a}. Based on this scheme and due to equations (\ref{conf}) and (\ref{H0}), we can write the Hamiltonian (\ref{ham}) as follows:
\begin{align}
H_M = H_0 + \Delta H,
\end{align}   
where 
\begin{align}
\label{deltaH}
&\Delta H = \frac{1}{2}\sum_{i = 1}^{M}\sum_{j=1}^{M}\Delta\Phi(q_{\alpha_i},q_{\alpha_j})\,,\\
&\Delta\Phi(q_{\alpha_i},q_{\alpha_j})=\Phi(q_{\alpha_i},q_{\alpha_j})-\phi_j(q_{\alpha_i})-
\phi_i(q_{\alpha_j}).
\end{align} 

Due to expressions (\ref{deltaH}), we can write the expression (\ref{conf}) in the form   
   \begin{align}
   \label{2.4}
   Q_N = \text{Sp}\Big\{\exp\left(-\beta H_0\right)\prod\limits_{i<j}\exp\left[-\beta \Delta\Phi(q_{\alpha_i},q_{\alpha_j})\right]\Big\}.
   \end{align}     

In order to expand expression (\ref{2.4}) to cumulants, we use modified Mayer functions as in \cite{Bokun18a} 
\begin{align}
f(q_{\alpha_i},q_{\alpha_j})=\exp\left[-\beta\Delta\Phi(q_{\alpha_i},q_{\alpha_j})\right] -1.
\end{align}

Expression (\ref{2.4}) can then be presented as the average over the reference distribution 
\begin{align}
\label{Q_N}
Q_N=Q_N^V\Big\langle\prod_{i<j}\big[1+f(q_{\alpha_i},q_{\alpha_j})\big]\Big\rangle_0\,\,,
\end{align}
where 
\begin{align}
\left\langle \ldots \right\rangle_0 &= \text{Sp}\big[\exp(-\beta H_0) \ldots \big],\\
Q_N^V &= \text{Sp}\big[\exp(-\beta H_0)\big].
\end{align}

Since equations (\ref{uu}) and (\ref{f10}) make it possible to take the averages of quantities of singlet and binary types, let us write expression (\ref{Q_N}) with the following terms of the expansion    
\begin{align}
\label{QN1}
Q_N = Q_N^V\bigg(1+\sum_{i<j}^{M}\sum_{\alpha_i,\alpha_j=0}^{1}f_{\alpha_i,\alpha_j}+\ldots \bigg),
\end{align}
where
\begin{align}
f_{\alpha_i,\alpha_j}=\int_{\omega_i}\int_{\omega_j}f(q_{\alpha_i},q_{\alpha_j})F_2^0(q_{\alpha_i},q_{\alpha_j})\rd q_{\alpha_i} \rd q_{\alpha_j}
\end{align}
and $F_2^0(q_{\alpha_i},q_{\alpha_j})$ being the pair distribution function (\ref{f23}) of the reference system. 

Expanding expression (\ref{QN1}) to cumulants, we obtain the free energy in the form of a functional
\begin{align}
\label{free}
F &= F_0 + \sum_{i=1}^{M}\sum_{j(i)}^{M}\sum_{\alpha_i,\alpha_j=0}^{1}f_{\alpha_i,\alpha_j}+ \ldots \,, \\
F & = \ln Q_N\,,\qquad F_0 = \ln Q_N^V\,.
\end{align} 

The right-hand side of equation~(\ref{free}) can be considered in a diagram representation \cite{Bokun00}. In order to derive equations determining the mean potentials, we use the condition
\begin{align}
\label{condi}
\sum_{\alpha_j=0}^{1}\int_{\omega_j} \rd q_{\alpha_j}f(q_{\alpha_i},q_{\alpha_j})F_2^0(q_{\alpha_i},q_{\alpha_j})=0\,,
\end{align}
which means that all the diagrams containing at least one free node are taken into account. In this case, the free energy functionals for the initial and the reference systems, therefore, coincide for the same distributions of the number of particles over the volume of the system.

We derive the equation for $\phi_j(q_{\alpha_i})$ by replacing equation (\ref{f10}) into equation (\ref{condi}) and due to equation (\ref{f22}). As a result, we have 
\begin{align}
\label{2.14}
\exp\left[-\beta\phi_j(q_{\alpha_i})\right]s_j(q_{\alpha_i})=
\sum_{\alpha_j=0}^{1}\frac{c_{\alpha_j}}{Q_{\alpha_j}}\int_{\omega_j}K(q_{\alpha_i},q_{\alpha_j})
s_i^{-1}(q_{\alpha_j})\exp\Big[-\beta\sum_{k\ne i,j}\phi_k(q_{\alpha_j})\Big] \rd q_{\alpha_j}\,,
\end{align}     
where
\begin{align}
\label{salr}
K(q_{\alpha_i},q_{\alpha_j})&=\exp\left[-\beta U_{\text{SL}}(q_{\alpha_i},q_{\alpha_j})\right],\\
U_{\text{SL}}(q_{\alpha_i},q_{\alpha_j})&=\Phi(q_{\alpha_i},q_{\alpha_j})+\sigma(q_{\alpha_i},q_{\alpha_j})\,. 
\end{align}

$U_{\text{SL}}$ is the potential consisting of short-range and screening pair potentials. This potential, in principle, can reproduce the short-range attractive and long-range repulsive (SALR) interaction, which is used to describe systems with competing interactions. 

The systems of equations (\ref{f23}), due to equations (\ref{H0}), (\ref{2.14}), (\ref{f22}), allow one to find the potentials $\phi_j(q_{\alpha_i})$ and the normalization functions $S_j(q_{\alpha_i})$, which can be used to find the singlet and the pair distribution functions. This makes it possible to determine thermodynamic properties of the system and calculate the free energy functional. As mentioned above, the latter can be found from the fact that in our case $F$ and $F_0$ coincide. 

Hence, in order to calculate the functional of the free energy of the initial system with the Hamil\-tonian~(\ref{ham}), it is sufficient to know the free energy of the reference system with the Hamiltonian~(\ref{H0}).

\section{The free energy of the reference system}

The configurational integral $Q_N^V$ can be found from the fact that the Hamiltonian (\ref{H0}) consists of pair interactions, and therefore
\begin{align}
\label{partial}
-\dfrac{\partial\ln Q_N^V}{\partial\beta}=\left\langle H_0\right\rangle_0 = \sum_{i=1}^{M}\sum_{\alpha_i = 0}^{1}
\left\langle U(q_{\alpha_i})\right\rangle_0+\frac{1}{2}\sum_{i=1}^{M}\sum_{j(i)}^{M}\sum_{\alpha_i,\alpha_j=0}^{1}\left\langle V(q_{\alpha_i},q_{\alpha_j})\right\rangle_0.
\end{align}  

Expression (\ref{partial}) contains the averages of the quantities of the singlet and binary types, which can be taken using the known functions (\ref{uu}), (\ref{f1}) and (\ref{f10}):
\begin{align}
\label{fs}
\sum_{\alpha_i = 0}^{1}
\left\langle U(q_{\alpha_i})\right\rangle_0 = \sum_{\alpha_i = 0}^{1}\int_{\omega_i}U(q_{\alpha_i})
F_1^0(q_{\alpha_i}) \rd q_{\alpha_i}.
\end{align}

Replacing expression (\ref{uu}) into equation (\ref{fs}), we have 
\begin{align}
\label{fs2}
\sum_{\alpha_i = 0}^{1}
\left\langle U(q_{\alpha_i})\right\rangle_0 = \sum_{\alpha_i = 0}^{1}\frac{c_{\alpha_i}}{Q_{\alpha_i}}\int_{\omega_i}U(q_{\alpha_i})\exp\left[-\beta U(q_{\alpha_i})\right]\rd q_{\alpha_i}.
\end{align}

Due to equation (\ref{f1}), we can rewrite this equation as follows:
\begin{align}
 \label{fs3}
 \sum_{\alpha_i = 0}^{1}
 \left\langle U(q_{\alpha_i})\right\rangle_0 =-\sum_{\alpha_i = 0}^{1}c_{\alpha_i}\dfrac{\partial\ln Q_{\alpha_i}}{\partial\beta}.
 \end{align}

Replacing equation (\ref{fs3}) into equation (\ref{partial}) and due to equation (\ref{f10}) we can write 
\begin{align}
\label{dp}
\dfrac{\partial\ln Q_N^V}{\partial\beta}=\dfrac{\partial}{\partial\beta}\sum_{i=1}^{M}\sum_{\alpha_i=0}^{1}\ln Q_{\alpha_i}^{c_{\alpha_i}}-\frac{1}{2}\sum_{i,j=1}^{M}\sum_{\alpha_i,\alpha_j}V(\alpha_i,\alpha_j)\,,
\end{align}
where
\begin{align}
V(\alpha_i,\alpha_j)=\int_{\omega_i}\int_{\omega_j}V(q_{\alpha_i},q_{\alpha_j})F_2^0(q_{\alpha_i},q_{\alpha_j})\rd q_{\alpha_i} \rd q_{\alpha_j}.
\end{align}

Integrating (\ref{dp}) from $0$ to $\beta$ and due to the relationship
\begin{align}
Q_N^V\Big\vert_{\beta = 0}=\omega^{M-N}\prod_{i=1}^{M}\prod_{\alpha_i=0}^{1}\frac{1}{c_{\alpha_i}}
\end{align}
we find
\begin{align}
\label{v1}
\ln Q_N^V=\sum_{i=1}^{M}\sum_{\alpha_i=0}^{1}\left[c_{\alpha_i}\ln\frac{Q_{\alpha_i}}
{c_{\alpha_i}}+V_{\alpha_i}^{(\beta)}\right],
\end{align} 
where

\begin{align}
\label{v2}
V_{\alpha_i}^{(\beta)}=\frac{1}{2}\int_{0}^{\beta} \rd\beta\sum_{j(i)}^{M}\sum_{\alpha_j=0}^{1}
\int_{\omega_i}\int_{\omega_j}V(q_{\alpha_i}q_{\alpha_j})F_2^0(q_{\alpha_i}q_{\alpha_j}) \rd q_{\alpha_i} \rd q_{\alpha_j}.
\end{align}

Due to equations (\ref{f23}) and (\ref{2.14}), the relationships (\ref{v1}) and (\ref{v2}) make it possible to find the free energy functional for a system with short-range and long-range interactions.

We will solve this problem for the case when the movement of molecules around lattice sites is neglected, which corresponds to the description of the crystal state of matter in the framework of the lattice approximation.

\section{Lattice approximation for the description of crystal and liquid states}
    
If we take into consideration only the states with particles occupying the sites (centers of cells) of the lattice, the kernels of equations (\ref{f23}) and (\ref{2.14}) depend only on the distance between the sites of the lattice and on the occupation numbers and become the functions of site variables $\alpha_i$ and $\alpha_j$. Consequently, the integral equations are reduced to algebraic equations. For instance, the equations of the system (\ref{f23}) take on the form 
\begin{align}
\label{4.1}
s_{j,\alpha_i}=\sum_{\alpha_j=0}^{1}s_{i,\alpha_j}^{-1}c_{\alpha_j}R_{\alpha_i,\alpha_j}\,,
\end{align}
where 
\begin{align}
R_{\alpha_i,\alpha_j}=\exp\big[-\beta\sigma(q_{\alpha_i}^*,q_{\alpha_j}^*)\big], 
\end{align}
with $q_{\alpha_i}^*$ and $q_{\alpha_i}^*$ denoting the coordinates of particles of the ``sorts'' $\alpha_i$ and $\alpha_j$ fixed in lattice sites $i$ and~$j$. Also, for the normalization functions we now use lower case notation: $S_j(q_{\alpha_i})\longrightarrow s_{j, \alpha_i}\,$.   

In contrast to the equations (\ref{f23}) and (\ref{2.14}), which are expressed in terms of coupled functions, in the case when the movement of particles near sites is neglected, the equation resulting from (\ref{f23}) turns out to be closed. This allows us to find the solution without referring to equation (\ref{2.14}). To this end, we write eguation~(\ref{4.1}) in the form 
\begin{align}
\label{4.3}
s_{j,0_i}&=\frac{c_{0_j}}{s_{i,0_j}}+\frac{c_{1_j}}{s_{i,1_j}}\,,\\
\label{4.4}
s_{j,1_i}&=\frac{c_{0_j}}{s_{i,0_j}}+\frac{c_{1_j}}{s_{i,1_j}}R_{ij}.
\end{align}      

Equations (\ref{4.3}) and (\ref{4.4}) take into account the absence of interaction between vacancies, i.e.,
\begin{align}
R_{0_i,0_j}=R_{0_1,1_j}=R_{1_i,0_j}=1.
\end{align}

We will consider the case of a homogeneous system. This means that $c_{1_j}=c_1$, $c_{0_j}=c_0$, and due to equations (\ref{4.3}) and (\ref{4.4}), we have 
\begin{align}
\label{4.6}
s_{j,1_i}=s_{i,1_j}=s_{ij}^{(1)},\\
s_{j,0_i}=s_{i,0_j}=s_{ij}^{(0)}.\nonumber
\end{align}

The system of equations (\ref{4.3}) and (\ref{4.4}) now takes on the form
\begin{align}
\label{4.7}
s_{ij}^{(0)}&=\frac{c_0}{s_{ij}^{(0)}}+\frac{c_1}{s_{ij}^{(1)}}\,,\\
s_{ij}^{(1)}&=\frac{c_0}{s_{ij}^{(0)}}+\frac{c_1}{s_{ij}^{(1)}}R_{ij}\,.\nonumber
\end{align}   

It is convenient to express the solution of equations (\ref{4.7}) in terms of the variables
\begin{align}
\label{4.8}
\eta_{ij}=\frac{s_{ij}^{(1)}}{s_{ij}^{(0)}}\,, \qquad K_{ij}=s_{ij}^{(0)}s_{ij}^{(1)},
\end{align}
which can be found from  equation~(\ref{4.7}) as follows:
\begin{align}
\label{4.9}
\eta_{ij}&=-\frac{c_1-c_0}{2c_0}+\sqrt{\left(\frac{c_1-c_0}{2c_0}\right)^2+\frac{c_1}{c_0}R_{ij}}\,\,,\\
\label{4.10}
K_{ij}&=\eta_{ij}c_0+c_1=c_0+\frac{c_1}{\eta_{ij}}R_{ij}\,,
\end{align}
and
\begin{align}
\label{4.11}
s_{ij}^{(1)}=\sqrt{K_{ij}\eta_{ij}}\,\,,\qquad s_{ij}^{(0)}=\sqrt{\frac{K_{ij}}{\eta_{ij}}}\,.
\end{align}

The relationships (\ref{4.9})--(\ref{4.11}) determine the normalization of the constant of the function (\ref{f10}), for which the following relationship holds true
\begin{align}
\label{4.12}
F_2^0(1i,1j)=c_1^2\frac{R_{ij}}{K_{ij}\eta_{ij}}\,.
\end{align}

Due to equations (\ref{4.9}) and (\ref{4.10}), the expression (\ref{4.12}) becomes
\begin{align}
\label{4.13}
F_2^0(1i,1j)=\frac{2c_1^2R_{ij}}{(c_0-c_1)+2c_1R_{ij}+\sqrt{(c_1-c_0)^2+4c_0c_1R_{ij}}}\,.
\end{align}

Based on equation (\ref{4.13}), it is possible to calculate the quantity (\ref{v2}), which determines the contribution of the long-range potential to the free energy. Equation (\ref{4.13}) tells us that in the lattice approximation this contribution does not depend on the short-range interaction. Mutual effect of short- and long-range interactions in this approximation is realized via the mean potentials defined by the system of equations~(\ref{2.14}), which in the lattice approximation can be written as follows:
\begin{align}
\label{4.14}
\epsilon_{j,\alpha_i} = \sum_{\alpha_j=0}^{1}\epsilon_{i,\alpha_j}^{-1}c_{\alpha_j}L_{\alpha_i,\alpha_j}\,,
\end{align} 
where
\begin{align}
\label{4.15}
\epsilon_{j,\alpha_i}&=\chi_{j,\alpha_i}s_{j,\alpha_i}\,,\qquad\chi_{j,\alpha_i}= \exp\big[-\beta\phi_k(q_{\alpha_i}^{*})\big]\,,\\
\label{4.16}
L_{\alpha_i,\alpha_j}&=R_{\alpha_i,\alpha_j}W_{\alpha_i,\alpha_j}\,,\qquad
W_{\alpha_i,\alpha_j}=\exp\big[-\beta\Phi(q_{\alpha_i}^{*},q_{\alpha_j}^{*})\big].
\end{align}

Equations (\ref{4.14}) have the same structure as equations (\ref{4.1}) which means that we can write the solution in a form similar to the relationships (\ref{4.8})--(\ref{4.11}). We denote the results of this solution with an asterisk:
\begin{align}
\label{4.17}
\eta_{ij}^{*}=\frac{\epsilon_{ij}^{(1)}}{\epsilon_{ij}^{(0)}}\,, \qquad K_{ij}^*=\epsilon_{ij}^{(0)}\epsilon_{ij}^{(1)},
\end{align}
which can be found from equation (\ref{4.7}) as follows:
\begin{align}
\label{4.18}
\eta_{ij}^*&=-\frac{c_1-c_0}{2c_0}+\sqrt{\left(\frac{c_1-c_0}{2c_0}\right)^2+\frac{c_1}{c_0}L_{ij}}\,,\\
\label{4.19}
K_{ij}^*&=\eta_{ij}c_0+c_1=c_0+\frac{c_1}{\eta_{ij}^*}L_{ij}\,,
\end{align}
and
\begin{align}
\label{4.20}
s_{ij}^{(1)}=\sqrt{K_{ij}^*\eta_{ij}^*}\,,\qquad s_{ij}^{(0)}=\sqrt{\frac{K_{ij}^*}{\eta_{ij}^*}}\,.
\end{align}

The solution (\ref{4.14}) makes it possible to find the part of the free energy (\ref{v1}) which is expressed in terms of the mean potentials. From expressions (\ref{4.15}) we have
\begin{align}
\label{4.21}
Q_{\alpha_i}=\omega\prod_{j(i)}\chi_{j,\alpha_i}=\omega\prod_{j(i)}\left(\frac{\epsilon_{j,\alpha_i}}{s_{j,\alpha_i}}\right).
\end{align}

Due to the relationships (\ref{4.20}) and (\ref{4.11}) we can write equation (\ref{4.21}) in the form
\begin{align}
\label{4.22}
Q_{1_i}=\prod_{j(i)}\left(\frac{K_{i,j}^*\eta_{i,j}^*}{K_{i,j}\eta_{i,j}}\right)^{1/2}\,,\quad
Q_{0_i}=\prod_{j(i)}\left(\frac{K_{i,j}^*\eta_{i,j}}{K_{i,j}\eta_{i,j}^*}\right)^{1/2}\,,
\end{align} 
which allows us to calculate the quantity (\ref{v1}). 

In contrast to the previous results for systems with short-range interactions \cite{Bokun00, Argyrakis01}, the free energy is determined to be a sequential procedure. The long-range contribution is taken into account through the normalization parameters $s_{ij}$, and then the short-range interactions are included through the mean potentials although the constitutive equations (\ref{4.9}), (\ref{4.18}) have the same structure with different kernels $R_{ij}$ and $L_{ij}$, correspondingly.

The lattice approximation, which has been considered here for the description of the crystal state, can also be applied to the description of the liquid state of matter. In the former case, we considered the states where the particles are located in the sites of the lattice. Now we should account for the fact that for a homogeneous medium the singlet distribution function, given by equation (\ref{f1}), is constant, and, therefore, the potentials $\phi_j(q_{\alpha_i})$ are also constant. Applying this condition to the equations (\ref{f23}) and (\ref{2.14}), we can once again reduce integral equations to algebraic ones. These algebraic equations determine the quantities $s_{j,\alpha_i}$ and $\chi_{j,\alpha_i}$ from equations (\ref{4.1}) and (\ref{4.14}), but instead of the kernels $R_{\alpha_i,\alpha_j}$ and $L_{\alpha_i,\alpha_j}$ we use the kernels $R_{\alpha_i,\alpha_j}^l$ and $L_{\alpha_i,\alpha_j}^l$ determined by the formulae
\begin{align}
\label{4.23}
R_{\alpha_i,\alpha_j}^l&=\frac{1}{\omega}\int_{\omega_j}\exp\big[-\beta\sigma(q_\alpha^*,q_{\alpha_j})\big] \rd q_{\alpha_j}\,,\\
\label{4.24}
L_{\alpha_i,\alpha_j}^l&=\frac{1}{\omega}\int_{\omega_j}\exp\big[-\beta\sigma(q_\alpha^*,q_{\alpha_j})-\beta\Phi(q_\alpha^*,q_{\alpha_j})\big]\rd q_{\alpha_j}.
\end{align}   

In summary, the equations (\ref{4.9})--(\ref{4.11}), (\ref{4.13}), (\ref{4.17})--(\ref{4.20}), (\ref{4.22}), (\ref{4.23}), (\ref{4.24}) allow one to calculate the functional of the free energy of a condensed system described by both short-range and long-range interactions. 

\section{The free energy and the chemical potential of a homogeneous system}

We first consider the contribution of the long-range interactions to the properties of the system. In the homogeneous case, in order to calculate the contribution from the Coulomb interaction, due to the electroneutrality condition one should consider the correlation function    
\begin{align}
\label{5.1}
h_2^0(q_{\alpha_i},q_{\alpha_j})=F_2^0(q_{\alpha_i},q_{\alpha_j})-F_1^0(q_{\alpha_i})F_1^0(q_{\alpha_j}).
\end{align}   

Using equation (\ref{4.13}), we can write equation (\ref{5.1}) in the form
\begin{align}
\label{5.2}
h_{ij}=\frac{2c^2R_{ij}}{2cR_{ij}+(1-2c)+\sqrt{(1-2c)^2+4c(1-c)R_{ij}}}-c^2,
\end{align}
where $c$ is the concentration of particles. 

Since in the equation (\ref{v2}) there is integration by the inverse temperature, for subsequent calculations it is convenient to introduce the reduced temperature $T_\text{b}$ characteristic of the system. As a result, we can introduce the ratio $\beta^*=T_\text{b}/T$. In accordance with \cite{Bokun18}, in terms of these reduced units we can write
\begin{align}
\label{5.3}
R_{ij}=\exp\left[-\beta\frac{r_\text{b}}{r_{ij}}\exp\left(-\sqrt{\beta}\nu r_{ij}\right)\right],
\end{align}   
where 
\begin{align}
r_\text{b}=\frac{e^2}{4\piup\varepsilon\varepsilon_0kT_\text{b}H}\,,
\end{align}
$e$ is the ion charge, $H$ is the lattice parameter, $\varepsilon$ is the permittivity, $\varepsilon_0$ is the dielectric constant, $k$ is the Boltzmann constant, $r_{ij}$ is the distance between the sites $i$ and $j$ in the units of the lattice parameter $H$,
\begin{align}
\label{5.5}
\nu=\frac{2}{H}\sqrt{\piup c(1-c)r_\text{b}}\,.
\end{align} 

We denote $T_\text{b}$ as the temperature, for which in $H$ units the Bjerrum length $r_\text{b}=1$.

Since the system is homogeneous and due to the fact that the left-hand side of equation (\ref{5.2}) depends only on the distance between the sites of the lattice, the quantity of interest (\ref{v2}) can be written as the sum
\begin{align}
\label{5.6}
V^{\beta}=\frac{1}{2}\sum_{k=1}^{k_\text{max}}z_k V_k\,,
\end{align}
where $z_k$ is the number of neighbors in the $k$-th coordination sphere, $V_k$ is the value of the Coulomb interaction potential averaged by the function (\ref{5.2}) at a distance $r_k$, which is the radius of the $k$-th coordination sphere. Since
\begin{align}
V(r_k)=\frac{r_\text{b}}{r_k}kT_\text{b}\,,
\end{align}
we have that in the equation (\ref{5.6}) the term
\begin{align}
\label{5.8}
V_k=\frac{r_\text{b}}{r_k}\int_{0}^{\beta^*}h_k \rd\beta^*,
\end{align} 
$h_k$ being the value of the function $h_{ij}$ at the distance $r_{ij}=r_k$.

Taking the derivative of equation (\ref{5.8}) with respect to the concentration, we can find the excess chemical potential due to the long-range interaction for each coordination sphere. In general, in accordance with the equation (\ref{v1}), the chemical potential of the system consists of three terms
\begin{align}
\mu=\mu^\text{id}+\mu^\text{sh}+\mu^\text{long}\,,
\end{align}  
where
\begin{align}
\beta\mu^\text{id}&=\ln\left(\frac{c}{1-c}\right),\\
\label{5.11}
\beta\mu^\text{sh}&=-z_1\left(\frac{\partial}{\partial c}\ln Q_{1_i}-\frac{\partial}{\partial c}\ln Q_{0_i}\right),\\
\label{5.12}
\beta\mu^\text{long}&=\sum_{k=1}^{k_\text{max}}z_k\frac{\rd \nu_k}{\rd c},
\end{align}
and $z_k$ is the number of neighbors in the $k$-th coordination sphere.

In the equation (\ref{5.11}), it is usually sufficient to consider only the first neighbors, whereas in the equation (\ref{5.12}) one should take into account several coordination spheres.

\section{The long- and short-range components of the chemical potential}

\begin{figure}[!b]
	\begin{center}
		\includegraphics [height=0.35\textwidth]  {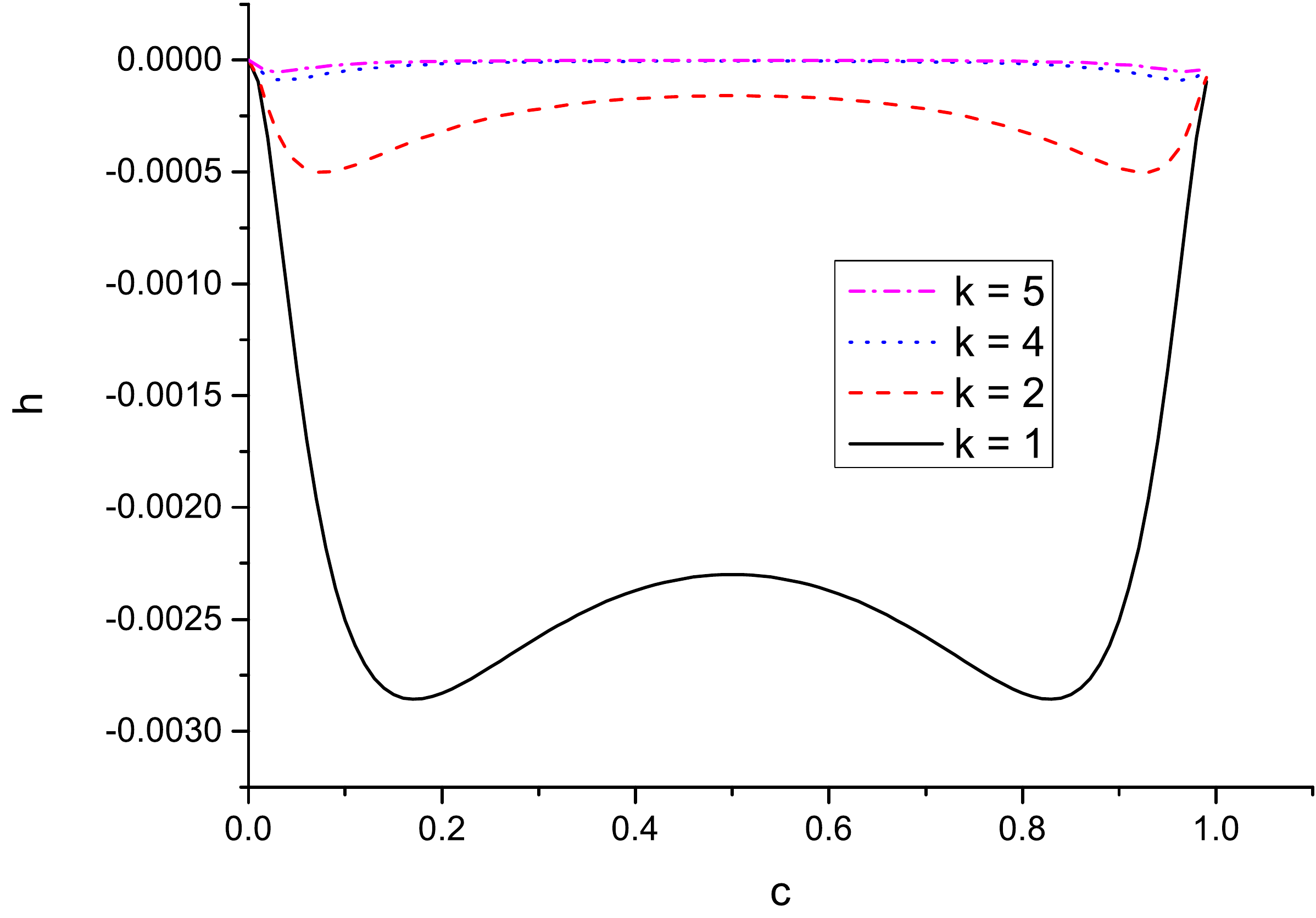}
		\includegraphics [height=0.35\textwidth]  {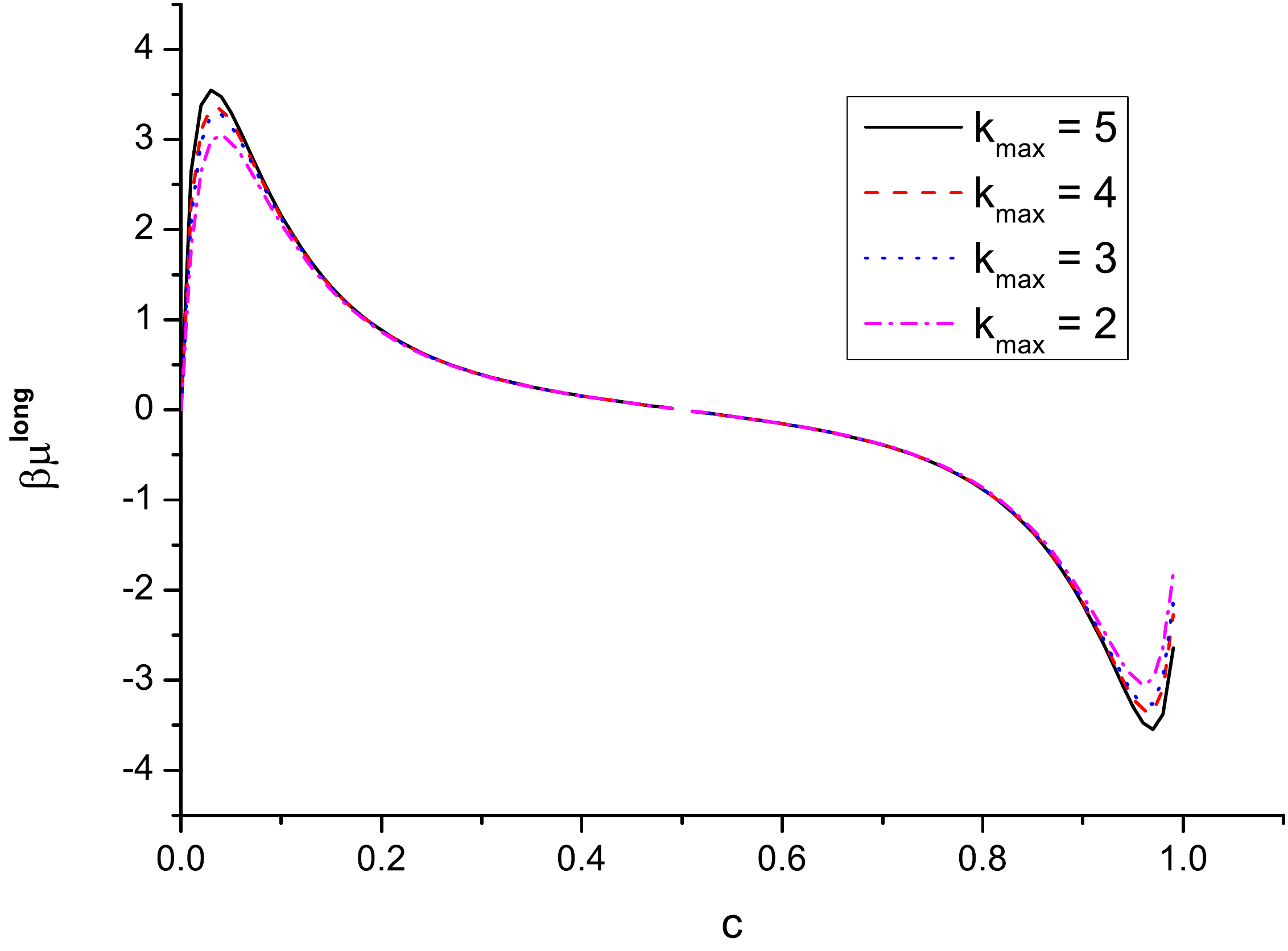}
		\caption{(Colour online) Left-hand panel: the value of the pair correlation function as a function of the concentration for different coordination spheres $k$. Right-hand panel: the long-range part of the chemical potential as a function of the concentration for different numbers of coordination spheres taken into account according to the equation (\ref{5.12}). The curves correspond to the value of $\beta = 10$.}
		\label{fig:Fig1}
	\end{center}
\end{figure}

According to the expression (\ref{5.12}), one needs to add up the contributions (\ref{5.8}) for a sufficient number of coordination spheres $k_\text{max}$ subject to the condition of $k$ approaching asymptotically the maximum value. We will, therefore, do the calculations for different values of $k_\text{max}$. As one can see from equations~(\ref{5.8}) and (\ref{5.2}), in order to find each term in the equation (\ref{5.12}), one should integrate the function (\ref{5.2}) over $\beta$ and differentiate it with respect to the concentration. We note that previously in \cite{Bokun18} this function was approximated by the expression (\ref{5.2}), which did not contain the denominator. The integration of this expression over $\beta$ led to the divergence at $c\rightarrow 1$ due to the dependence (\ref{5.5}). This resulted in the loss of thermodynamic stability even at small concentrations, and the symmetry of the chemical potential as a function of the concentration was broken. Hence, the description considered in \cite{Bokun18} is justified only for the case when $c\ll1$. Due to this, we will analyse the dependence of the correlation function (\ref{5.2}), which we use instead of the expression (\ref{5.3}). The numerical analysis of the expression (\ref{5.3}) has shown that imposing the normalization conditions on the correlation functions corrects the above mentioned flaw of the previous approach. Notably, analytical expansion of equation (\ref{5.3}) on the concentration shows that at $c\rightarrow 0 $ the function $h_k \sim c^2$, while at $c\rightarrow 1 $ we have $h_k \sim (1-c)^2$. In addition, for all $k$, the function $h_k$ has the correct symmetry with respect to $c=0.5$, which we can see on the left-hand panel of figure~\ref{fig:Fig1}.    

In order to find the contribution of the long-range interaction to the chemical potential, we swap the operations of integration and differentiation in equations (\ref{5.8}) and (\ref{5.12}). As a result, for the case of a simple cubic lattice, we obtain the chemical potential as a function of the concentration $c$, the inverse temperature $\beta$, and the number of coordination spheres $k_\text{max}$ taken into account. From the curves of the chemical potential $\mu^\text{long}(k_\text{max},c,\beta)$ (right-hand panel of figure~\ref{fig:Fig1}) as a function of the concentration at $\beta = 10$ and $k_\text{max}=4$ and $5$, we can see that it is sufficient to take into account only five coordination spheres.

The phase transition in the reference system determines the dependence of the chemical potential of the reference system on the concentration
\begin{align}
\label{6.1}
\beta\mu_\text{b}=\ln\frac{c}{1-c}+\beta\mu^\text{long}.
\end{align}     

\begin{figure}[!t]
	\begin{center}
		\includegraphics [height=0.35\textwidth]  {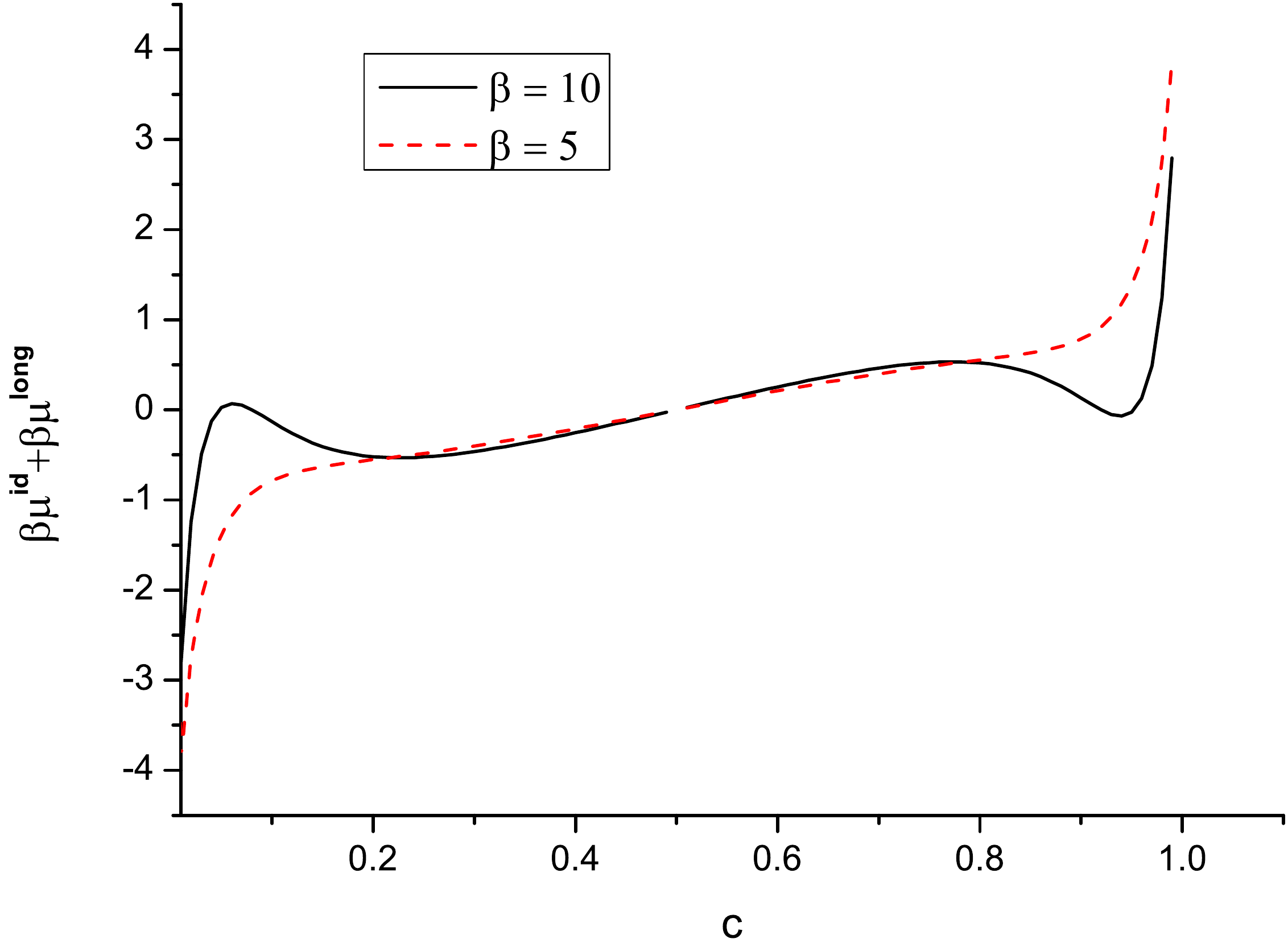}
		\includegraphics [height=0.35\textwidth]  {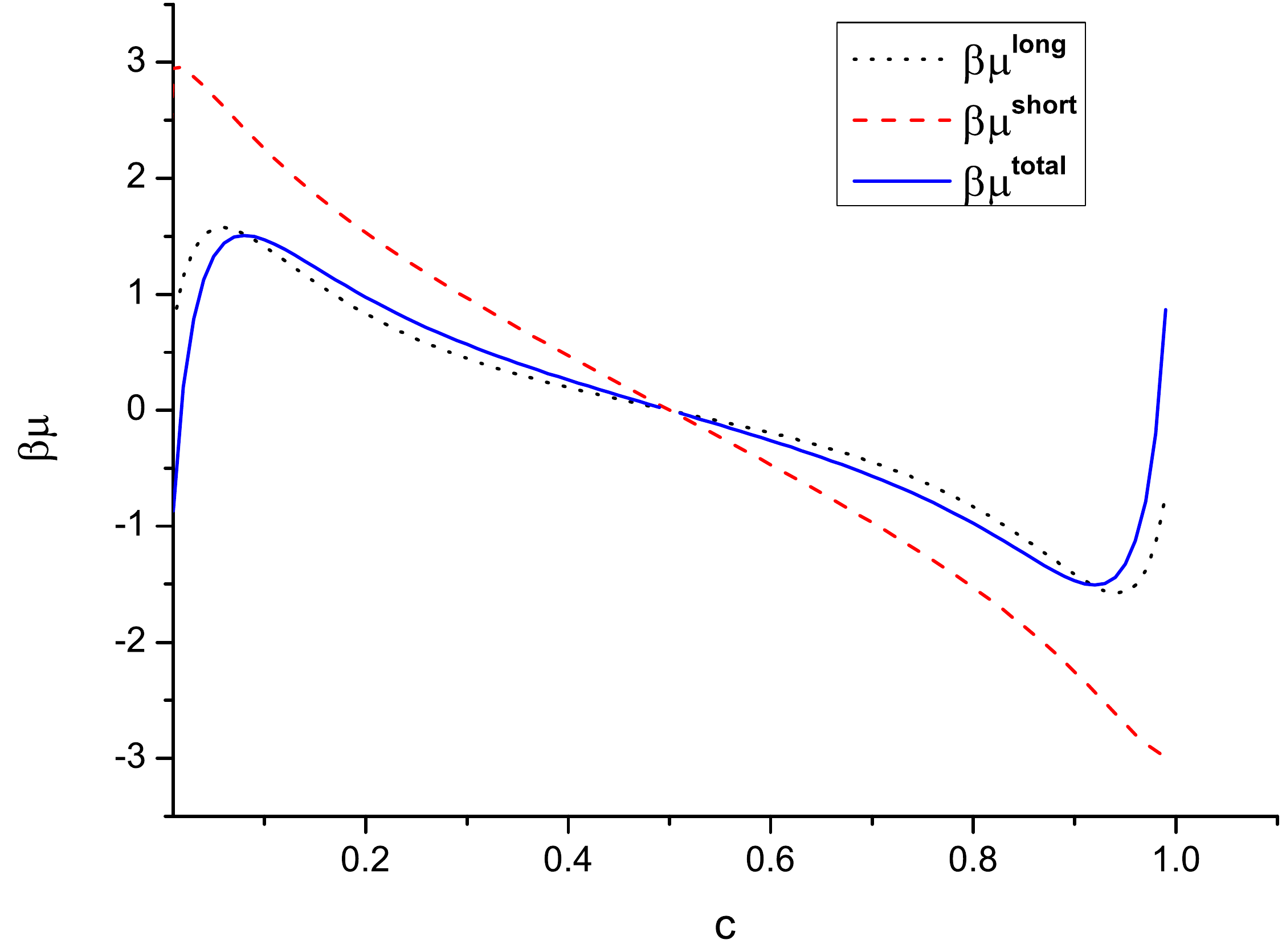}
		\caption{(Colour online) Left-hand panel: the chemical potential as the sum of ideal and long-range contributions given by the equation~(\ref{6.1}). Right-hand panel: short-range and long-range contributions to the chemical potential and their sum as functions of the concentration for $\beta J=-1$.}
		\label{fig:Fig2}
	\end{center}
\end{figure}

This dependence is shown in figure~\ref{fig:Fig2} (left-hand panel) for the temperatures above and below the critical temperature (for $\beta = 5$ and $\beta = 10$, respectively). It shows the possibility of a phase transition, the parameters of which are determined by the concentration such that $\mu = 0$.

The contribution from the short-range interaction can be found from the formula (\ref{5.11}). We take into account only the interaction between the nearest neighbors, which is characterized by the parameter $J$. In figure~\ref{fig:Fig2} (right-hand panel) we compare $\mu^\text{sh}(c,\beta)$ and $\mu^\text{long}(c,\beta)$ and show the total chemical potential as functions of the concentration. 

The figures presented describe the contributions of short-range and long-range interactions to equilibrium thermodynamic properties of high temperature conductors.   
 
Based on the results for the correlation function $F_{00}$ calculated from the formulae (\ref{f23}), (\ref{4.20}) and the compressibility, we can find the chemical diffusion coefficient as \cite{Argyrakis01, Zhdanov85}
\begin{align}
\label{6.2}
D = WF_{00}\left(\frac{\partial\beta\mu_\text{b}}{\partial c}\right),
\end{align} 
where $W$ is the frequency of particle oscillations around the lattice site, $\mu_\text{b}$ is the chemical potential of the reference system (\ref{6.1}), and $F_{00}$ is the probability that two neighbor sites are vacant. The latter quantity can be found as follows:
\begin{align}
\label{6.3}
F_{00}=F(0_i,0_j)=\frac{1}{\big[s_{ij}^{(0)}\big]^2}=\frac{\eta_{ij}}{K_{ij}}(1-c)^2.
\end{align}  

\begin{figure}[!t]
	\begin{center}		
		\includegraphics [height=0.35\textwidth]  {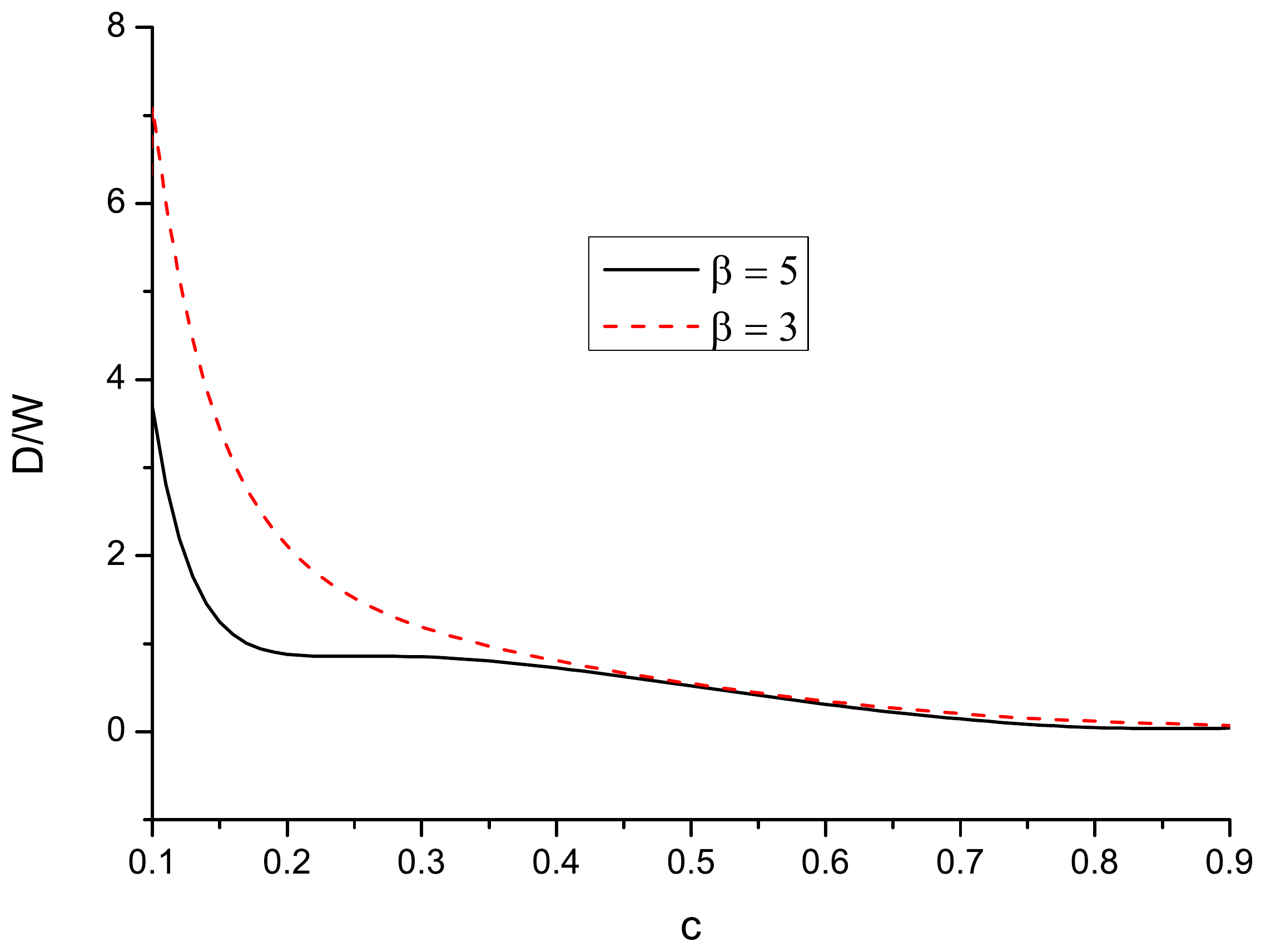}
		\caption{(Colour online) The reduced diffusion coefficient as a function of the concentration given by the equations (\ref{6.2}) and (\ref{6.3}).}
	\label{fig:Fig3}
	\end{center}
\end{figure}

In figure~\ref{fig:Fig3} we present the reduced transport characteristic $D_1=D/W$ as a function of the concentration. The diffusion coefficient strongly decreases with increasing the concentration and temperature.

\section{Conclusions}

In this paper we have proposed a statistical description of properties of solid electrolytes, for which the conductivity is caused by thermally activated ion mobility in the neutralizing background of immovable counterions. We have employed a scheme of statistical lattice description of condensed systems in order to model the migration of ions on the sites of their sub-lattices. In order to take into account both short-range and long-range interactions in a one-component system of charged particles, we have developed a joint approach combining the application of short-range mean potentials and the collective variables method that describes long-range interactions.

The approach is based on the expansion of the partition function on renormalized Mayer functions using mean potentials. As opposed to previous methods, the reference system includes not only mean potentials but also Coulomb interaction. In order to calculate the cumulants averaged over the reference distribution, we use the expression for the pair distribution function found by the collective variables method. The formula for the pair distribution function has been corrected to satisfy the normalization condition for the first terms of the expansion. This allowed us to calculate thermodynamic and diffusion properties of the considered materials in the entire region of thermodynamic parameters.

\section*{Acknowledgments}

This project has received funding from the European Union's Horizon 2020 research and innovation programme under the Marie Sk\l{}odowska-Curie grant agreement No 734276.

%\subsection{Ukrainian part}
%\label{ua-part}
%
%If you are not familiar with Ukrainian language, just uncomment
%two lines before \verb|\ukrainianpart| in the template. The
%necessary translation will be made by the Editorial Office.

%\bibliographystyle{cmpj}
%\bibliography{cmpjxampl}

%
%% If you have problems with typesetting in ukrainian uncomment lines below.
%
%  \lastpage
%  \end{document}

%\ukrainianpart
%
%\title{Повна назва: Зразок статті та поради авторам}
%\author{А.В. Тор\refaddr{label1,label2}, Б.В. Тор\refaddr{label2}}
%\addresses{
%\addr{label1} Університет ім. Орнштейна, Софтленд, 10041 Цельсій, вул. Реін, 1
%\addr{label2} Інститут ім. Церніке, Солідшир, 20451 Фаренгейт, пр. Рівер, 2
%}
%%
%%% якщо автор є один або автори є з однієї установи:
%%
%%  \author{1й Автор, 2й Автор, \ldots}
%%  \address{Інститут\ldots}
%%
%%%

\ukrainianpart

\title{Вклади від короткосяжних і далекосяжних взаємодій у рівноважні і транспортні властивості твердих електролітів}
\author{Г. Бокун\refaddr{label1}, І. Кравців\refaddr{label2}, М. Головко\refaddr{label2}, В. Віхренко\refaddr{label1}, Д. ді Капріо\refaddr{label3}}
\addresses{
	\addr{label1} Бiлоруський державний технологiчний унiверситет, вул. Свєрдлова, 13a, 220006 Мiнськ, Бiлорусь 
	\addr{label2} Iнститут фiзики конденсованих систем НАН України, вул. Свєнцiцького, 1, 79011 Львiв, Україна
	\addr{label3} Дослiдницький унiверситет науки та лiтератури Парижу, ChimieParisTech — CNRS, Iнститут хiмiчних
	дослiджень Парижу, Париж, Францiя
}
\makeukrtitle

\begin{abstract}
\tolerance=3000%
	В рамках комбінованої теорії, яка враховує далекосяжні та короткосяжні взаємодіЇ, описуються конденсовані іонні системи. Короткосяжна взаємодія виражається через середні потенціали, а далекосяжна взаємодія розглядається у формі екранованих потенціалів. Виходячи з умови оптимальної узгодженості розглядуваної системи із системою відліку, для цих потенціалів отримано систему інтегральних рівнянь. На відміну від опису середовища із короткосяжними взаємодіями, у даній роботі система відліку, окрім поля середніх потенціалів, включає також кулонівську взаємодію між частинками. Розглядається односортна система іонів у компенсуючому полі нерухомих іонів з протилежним зарядом. Модель може бути застосовано для опису твердих іонних провідників. Для вивчення руху катіонів на вузлах підгратки, використовується граткове наближення теорії для розрахунку парної функції розподілу. На основі методу колективних змінних пропонується методика, яка покращує вихідний вираз для цієї функції. Зокрема, доданки, які нехтуються при розвиненні цієї функції, апроксимуються одночастинковими доданками, які забезпечують умову нормування функцій розподілу. Таким чином, застосовність теорії узагальнено на ширшу область термодинамічних параметрів. Розраховано хімічний потенціал і коефіцієнт дифузії, які дозволяють описувати характерні для системи фазові переходи. 
	
\keywords керамічні провідники, середні потенціали, граткове наближення, метод колективних змінних, парна функція розподілу, хімічний потенціал
\end{abstract}

\end{document}